\documentclass[pre,
  aps,
  a4paper,
  english,
  showpacs,
  showkeys,
  reprint,
  twocolumn,
  superscriptaddress]{revtex4-1}
\usepackage[T1]{fontenc}
\usepackage[utf8]{inputenc}
\usepackage{amsmath,amsthm,amssymb,graphicx,subfigure,refstyle}
\usepackage{bm}
\usepackage{array,natbib,caption,amsfonts}
\usepackage{mathtools}
 \usepackage{lipsum}
\usepackage{capt-of,verbatim,url}
\usepackage[normalem]{ulem}
\usepackage{soul}
\usepackage[
  citecolor=blue,
  colorlinks,
  linkcolor=blue,
  urlcolor=blue,
]{hyperref}
\usepackage[dvipsnames]{xcolor}
\usepackage{hyperref}

\begin{document}
\newpage
\title{{\color{black}Where the Li\'enard--Levinson--Smith (LLS) theorem cannot be applied for a generalised Li\'enard system}}

\author{Sandip Saha}
\email{sandipsaha@bose.res.in} 
\affiliation{S. N. Bose National Centre for Basic Sciences, Block-JD, Sector-III, Salt Lake, Kolkata-700106, India}
\author{Gautam Gangopadhyay}
\email{gautam@bose.res.in} 
\affiliation{S. N. Bose National Centre for Basic Sciences, Block-JD, Sector-III, Salt Lake, Kolkata-700106, India}
\begin{abstract}
We have examined a class of Li\'enard--Levinson--Smith (LLS) system having a stable limit cycle which demonstrates the case {\color{black}where the LLS theorem cannot be applied}. The problem has been partly raised in a recent communication by Saha et al.~\cite{Tri-rhythmic} (last para of sec 4.2.2). Here we have provided a physical approach to address this problem using the concept of energy consumption per cycle. We have elaborated the idea through proper demonstration by considering a generalized model system. Such issues have potential utility in nonlinear vibration control.
\end{abstract}


\maketitle
It is the extraordinary perception of  Lord Rayleigh~\cite{rayleigh,strogatz,jordan} about the rhythmicity and quality of  tones  who first introduced nonlinear position dependent damping force to understand the self-oscillation or limit cycle as a connected exposition of the theory of nonlinear processes in dissipation and maintenance  of vibrational energy through proper shape and size of the musical instruments. 
As a general ground of vibration beyond  the sounds of stretched strings, bars, membranes and plates such subjects as ocean tides, not to speak of optics, and literally extended to any cyclic events where such novelty of treatment and results are followed with detailed consideration~\cite{strogatz,jordan,mickens}.
In open systems a limit cycle plays an important role as a feedback loop in dynamics in various kind of physical, chemical and biological processes, such as van der Pol oscillator~\cite{strogatz,jordan,mickens,Tri-rhythmic,limiso,len3.5}, Glycolytic oscillator (Selkov model)~~\cite{limiso,goldbook,glyscott,glysel1,glysel2,gly2,epstein}, Belousov--Zhabotinsky reaction~\cite{epstein}, Brusselator model for oscillatory chemical reactions~\cite{len4,limiso,epstein} and Circadian oscillator~\cite{goldbook,murray2,epstein,cir1,cir2} are some of the major examples. The variants of van der Pol oscillator  for physical circuits whereas circadian oscillator for biological rhythms~\cite{goldbook,murray2,cir1,cir2}, are prototypical testing grounds for  isolated closed trajectories  where the origin of such  competition between instability and damping can be investigated.

Li\'enard system~\cite{lienard,strogatz,jordan,nayfeh,mickens,len4,limiso} holds an important place in the theory of dynamical systems. It is basically a generalization of damped linear equation~\cite{strogatz,jordan,len3.5} with the coefficient of damping is replaced with position dependent damping coefficient. The general form of Li\'enard equation is, 
\begin{equation}
\ddot{x} + f(x)\dot{x} + g(x) = 0,
\end{equation}
where $f(x),\,g(x)$ are nonlinear 
functions and overhead dots represent the derivatives with respect to time. This system has been studied in great detail in Ref.~\cite{strogatz,jordan,len4,powerlaw}. One of the main aspect of Li\'enard system is the existence of limit cycles~\cite{rayleigh,strogatz,jordan,nayfeh} 
where the theorem entails conditions which primarily requires that $g(x)$ should be an odd analytic function and $f(x)$ should be positive in the neighbourhood of the origin for the existence of limit cycle. 

Li\'enard--Levinson--Smith (LLS)~\cite{lienard,mickens,len4,limiso,levinson1,levinson2}, proposed a more general form with the damping coefficient function depending on position as well as momentum of particle of the form~\cite{Tri-rhythmic},
\begin{equation}
\ddot{x} + F(x,\dot{x})\dot{x} + G(x) = 0,
\end{equation}
 where $F,\,G$ are arbitrary analytical functions. The casting of $LLS$ system from an arbitrary autonomous 2D kinetic flow equation is given in~\cite{len4}. The condition for the existence of a locally stable limit cycle is given in Ref.~\cite{mickens,levinson1,levinson2}(Appendix \ref{sec:Conditions}), where apart from the even and odd properties of $F$ and $G$, the condition $G.16$ of Ref.~\cite{mickens} (pp. 283, see Appendix \ref{sec:Conditions}), given as, $$C3:\quad F(0,0)<0,$$ also plays an important role.

\emph{The conditions in $LLS$ system (Ref.~\cite{mickens}; pp. 283) are similar to the Li\'enard theorem (Ref.~\cite{strogatz}; pp. 210), however, with the velocity dependence of the damping coefficient, the $C3$ condition is an exception which {\color{black}cannot be applied for a generalised Li\'enard system} and is going to be the focal point of this report.}
{\color{black}The point basically reduces to establish the condition, $F(0,0) < 0$ should be relaxed to $F(0,0) \le 0$, which we have performed here through a physical approach, by using  an approximate analytical tool (K-B averaging method) as well as a direct computational approach, considering a class of model systems.}

To understand the significance of condition $C3$, consider the van der Pol oscillator, 
\begin{equation}
\ddot{x} + \epsilon(x^2 - 1)\dot{x} + \omega^2 x =0, \quad \epsilon>0.
\end{equation}

Clearly, the condition for the existence of limit cycle, $f(0)(=-\epsilon<0)$ is satisfied and it is well known that a limit cycle exists for the same condition. Next, consider Rayleigh equation~\cite{strogatz,jordan,rayleigh,dsrrayleigh} as an example for $LLS$ system, 
\begin{equation}
\ddot{x}+\epsilon(\dot{x}^2 - 1)\dot{x}+x=0,\,\,\epsilon>0.
\end{equation}

Rayleigh oscillator equation models oscillation of a violin string and it is known that this system has a limit cycle which satisfies the condition $C3$, $F(0,0)(=-\epsilon<0)$. It denotes the instability of the origin which results in the gain of energy by the system only to be compensated by the damping as soon as $F(x,\dot{x})$ changes sign. Further, consider the Glycolytic oscillator of the $LLS$ form~\cite{limiso}, 
\begin{align}
\ddot{\xi}+\left[(1+a+3 b^2)-2 b \xi-2bk-3 b \dot{\xi} + \xi\dot{\xi} + k\dot{\xi} + \dot{\xi}^2 \right] \dot{\xi} \nonumber \\ 
+(a+b^2)\xi=0,
\end{align}
where $a,b>0$ and $k=b+\frac{b}{a+b^2}$. The system has a unique stable limit cycle for $F(0,0)<0$. 

\begin{figure}[h]
\includegraphics[width=\linewidth]{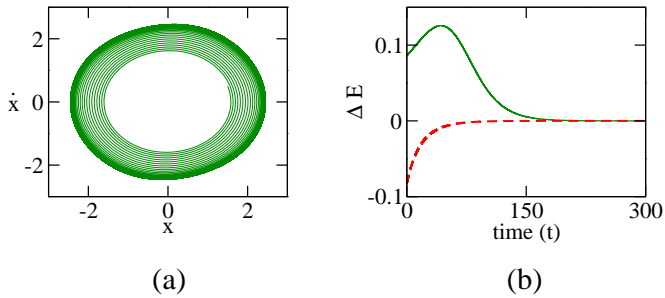}
\caption{(a) Phase portrait of Eq.~\ref{eq:cs1} for $\epsilon=0.01$ using the tool of approximate analytical solution by K-B method (b) shows the variation of energy change with respect to time for two different initial conditions where dashed (red) is for the outside initial condition and continuous (green) is for the initial condition lying inside the cycle. The energy change converges to null line denoting the existence of limit cycle. }
\label{fig:plot}
\end{figure}

Even though the above mentioned examples show the significance of the condition $C3$, it {\color{black}cannot be applied for a particular type of system~\cite{mickens2003}}. In~\cite{mickens2003}, Mickens considered the system
\begin{equation}
\ddot{x} + \epsilon(x^2 - 1)\dot{x}^{1/3} + x = 0
\end{equation}
and argued that the structural form of the differential equation occurring in the $LLS$ theorem cannot be applied to it. The argument rely on the fact that $F(0,0)$ blows up at origin and cannot form a valid condition for $LLS$. However, this condition fails even for finite values also. This claim is demonstrated in the following case study. Consider the system,
\begin{equation}
\ddot{x} + \epsilon \lbrace(x^2 - 1)\dot{x}^2\rbrace\dot{x} + x = 0,\quad 0<\epsilon \ll 1. \label{eq:cs1}
\end{equation}
Here, the damping coefficient function, $F(x,\dot{x})=\epsilon(x^2 - 1)\dot{x}^2$ is not satisfying the condition $F(0,0)<0$ as $F(0,0)=0$ although it has a unique stable limit cycle (Appendix \ref{sec:ProofUniqueSLC}) which we have verified numerically and the corresponding phase portrait is given in figure~\ref{fig:plot}(a).

The linear stability analysis for the system in Eq.~\ref{eq:cs1}  fails as the corresponding eigenvalues are purely imaginary, which is a condition for the center, however, the numerical simulation shows otherwise. We know that for a center solution~\cite{calogero}, which gives unique cycle for each initial condition, the energy change with time is zero from  initial time, however, for limit a cycle two separate initial conditions shall approach zero energy change after a finite time~\cite{powerlaw}.
 
To inspect the kind of solution of Eq.~\ref{eq:cs1}, we  consider the energy as that of a conservative system, and calculated the change in energy (see Appendix \ref{sec:EnergyChange}) as function of time i.e.$\Delta E=2 \, \pi \, \overline{r} \, \dot{\overline{r}}$ by using K-B averaging method~\cite{jordan,mickens} where $x(t) \approx r(t) \cos(t+\phi), r(t) \approx \overline{r}+O(\epsilon)$, is plotted in figure~\ref{fig:plot}(b). From the plot, one can find that the energy difference over a time period converges to zero value, which demonstrates the existence of the cycle after a  transient period. In the plot it is shown that the phase space trajectories for two different initial conditions approach the null energy lines. For a limit cycle the net change of system energy over a complete cycle is zero, the system energy change approaches this zero value, or the cycle by gaining or releasing the energy depending on whether the initial points are inside or outside of the cy
 cle, respectively. 

\begin{figure}[h]
\begin{center}
\includegraphics[width=\linewidth]{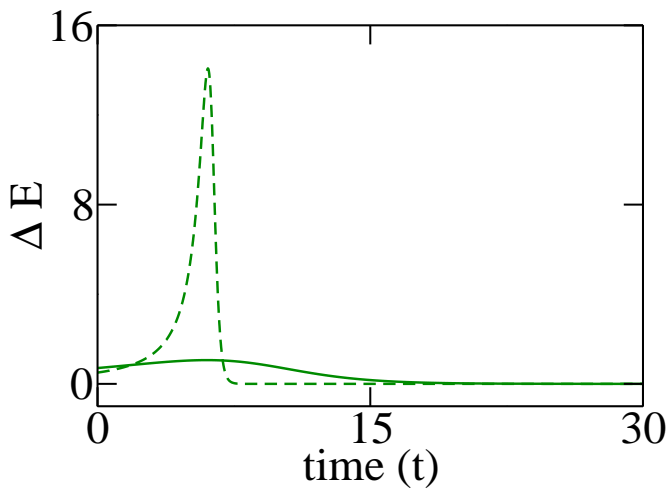}
\caption{Variation of energy change with respect to time for the proposed general system given by Eq.~\ref{eq:genLC} for the cases \{$a=1,\,m=1,\,n=2$\} (continuous) and \{$a=1,\,m=1,\,n=3$\} (dahsed) denoting the existence of limit cycle. }
\label{fig:plot2}
\end{center}
\end{figure}

This study shows that condition $F(0,0)<0$ for $LLS$ equation is {\color{black} failed to satisfy for a limit cycle system} and it could be revised in preference to a more general condition which incorporates such cases. Such equations could be generalised in the form
 \begin{equation}
 \ddot{x} + \epsilon \lbrace(x^{2m}-a^2)\dot{x}^{2n}\rbrace\dot{x} + x = 0, \label{eq:genLC}
 \end{equation}
where $m,\,n\in \mathbb{Z^+}$, $a\in \mathbb{R}-\lbrace 0 \rbrace$. Similar types of generalised systems has been analysed by Kovacic et al.~\cite{Kovacic2011, Kovacic2012} with $n=0$ and $\frac{1}{2}$. The first case shows the dynamics of a van der Pol like oscillator~\cite{Kovacic2012} and the second one provides an isochronous motion~\cite{calogero} where Chiellini integrability can be observed~\cite{Pandey2017}. In this study, we have considered the cases $n \ge 1$. Figure~\ref{fig:plot2} shows variation of energy change with respect to time for the proposed general system given by Eq.~\ref{eq:genLC} for the cases \{$a=1,\,m=1,\,n=2$\} (continuous) and \{$a=1,\,m=1,\,n=3$\} (dahsed) denoting the existence of limit cycle. It is also noted that with  increasing power of the velocity (for increasing $n$) the energy change falls more sharply with an elevated peak and for higher values of $m$ the peak is lowered with narrower width.

In summary, we have pointed out that the condition $C3$ is {\color{black} failed to satisfy for a limit cycle system} and demonstrate the fact using Eq.~\ref{eq:cs1} as the case study. The existence of limit cycle is established using the energy argument. Furthermore we have also provided a class of models where the condition fails. In the numerical examples we have shown that the influence of the higher integer power of the damping force not only decreases the time to reach the steady state but also drastically increases the energy change in the system which can be immensely useful in understanding the controlled tuning of sound vibration. By keeping the constraints of condition $C3$ remaining the same, one can also design and develop more general models which may have applications in nonlinear vibration control, network modelling, circuit design and related areas. 


\section*{Acknowledgment}
Sandip Saha acknowledges RGNF, UGC, India for the partial financial support. We are thankful to Prof. Deb Shankar Ray, Prof. Partha Guha and Dr. Sagar Chakraborty for some fruitful discussions. SS is grateful to Ankan Pandey for a lot of help during the writing of the initial version of the manuscript. SS is also thankful to  Mrs. Nibedita Konar for various academic related support while finalising this manuscript. 

\appendix
{
\section{Li\'enard--Levinson--Smith (LLS) theorem}
\label{sec:Conditions}
For a {\color{black}considered generalised form of} system{\color{black}, i.e.,} $\ddot{x}+F(x,\dot{x}) \dot{x}+G(x)=0$ ({\color{black} with} the arbitrary analytic functions $F$ and $G$){\color{black}, the existence of at least one limit cycle under certain conditions} (Ref.~\cite{mickens}; pp. 283):
\begin{enumerate}
\item[C1.] $xG(x)>0,\quad for\, |x|>0$
\item[C2.] $\int_0^\infty G(x) dx=\int_0^{-\infty} G(x)dx = \infty$
\item[C3.] $F(0,0)<0$
\item[C4.] $\exists\,\, x_0>0\,\,s.t. \quad F(x,\dot{x})\geq 0, \quad |x|\geq x_0$
\item[C5.] $\exists,\,\ M>0,\,\,s.t. \quad F(x,\dot{x})\geq -M, \quad |x|\leq x_0$
\item[C6.] $\exists\,\, x_1>x_0 \quad s.t. \int_{x_0}^{x_1} F(x,\dot{x}) dx \geq 10 M x_0${\color{black}, where $\dot{x}>0$ is an arbitrary decreasing positive function of $x$}.
\end{enumerate}
{\color{black}Under these conditions, $\exists$ at least one limit cycle of the considered equation.}
\section{Derivation of the change in energy ($\Delta E$)}
\label{sec:EnergyChange}

Consider the LLS system,
\begin{eqnarray}
\ddot{x}+F\left(x,\dot{x}\right)\dot{x}+ G(x)=0;~x=x(t),~G(x)= x.
\end{eqnarray}
The above expression can be written in the form of a weakly nonlinear oscillator as
\begin{eqnarray}
\ddot{x}+\epsilon h(x,\dot{x})+x=0,
\label{eq:bkform}
\end{eqnarray}
where $\epsilon~(0 < \epsilon \ll 1)$ is the nonlinearity control parameter and $h(x,\dot{x}) (=\frac{1}{\epsilon}F\left(x,\dot{x}\right)\dot{x})$ contains nonlinear damping terms.

To apply K-B perturbative method, let us choose, $x(t)\approx r(t)~\cos (t+\phi(t))$ then we have $r(t) \approx \sqrt{x^2 +\dot{x}^2}$ and $\phi(t) \approx -t+ ArcTan(- \frac{\dot{x}}{x})$, where $r$ and $\phi$ are the amplitude and phase, respectively. Then one can obtain $\dot{r} \approx \epsilon~h~\sin(t+\phi(t))$ and $\dot{\phi} \approx \frac{\epsilon~h}{r} \cos(t+\phi(t))$ i.e. the time derivative of  amplitude and phase are of $O(\epsilon)$. After taking a running  average~\cite{strogatz,mickens,powerlaw} of a time dependent function $U$ defined as, $\overline{U}(t) = \frac{1}{2 \pi} \int_{0}^{2\pi} U(s) ds$, one finds,
\begin{eqnarray}
\dot{\overline{r}} & \approx & \langle \epsilon~h~\sin (t+\phi(t)) \rangle_t = \varphi_1 (\overline{r},\overline{\phi}),\nonumber\\
\dot{\overline{\phi}} & \approx & \langle \frac{\epsilon ~h~}{r(t)} \cos (t+\phi(t))\rangle_t = \varphi_2 (\overline{r},\overline{\phi}).
\label{eq:ampphdynamics}
\end{eqnarray}
The functions $\varphi_1$ and $\varphi_2$ can be obtained from the explicit form of $h$ i.e., $F$ for the particular cases. Since $\dot{r}(t)$ and $\dot{\phi}(t)$ are of $O(\epsilon)$ then one can set the perturbation on $r(t)$ and $\phi(t)$ over one cycle as, $r(t) =\overline{r}+O(\epsilon)$ and  $\phi(t) =\overline{\phi}+O(\epsilon)$.

Therefore, we can calculate the approximate solution (i.e., $x(t) = \overline{r} \cos (t+\overline{\phi})+O(\epsilon)$) of Eq.~\ref{eq:bkform} by solving the above coupled amplitude-phase dynamics i.e., Eq.~\ref{eq:ampphdynamics}.

As $\epsilon$ is taken very small, one can define the system's approximate energy, as, $E \approx \frac{1}{2}(x^2+\dot{x}^2)$. Then the change or consumption of energy per cycle can be calculated as,
\begin{align}
\Delta E &= \int_{0}^{T} \frac{d E}{dt} dt = \int_{0}^{2 \pi +O(\epsilon)} \frac{d E}{dt} dt = 2 \pi \overline{r}~ \dot{\overline{r}} = \frac{d}{dt}(\pi \overline{r}^2),
\end{align}
where $O(\epsilon^2)$ are neglected.
{\color{black}
\section{Proof of unique stable limit cycle for system~\ref{eq:cs1}}
\label{sec:ProofUniqueSLC}
Using K-B approach (described in Appendix~\ref{sec:EnergyChange}) for system~\ref{eq:cs1}, we can find the amplitude-phase dynamics (cf.~\ref{eq:ampphdynamics}) as,
\begin{subequations}
\begin{eqnarray}
\dot{\overline{r}} &=& -\frac{1}{16}~\epsilon~\overline{r}^3 \left(\overline{r}^2-6\right), \label{eq:amplitudesys7}\\
\dot{\overline{\phi}} &=& 0.
\end{eqnarray}
\end{subequations}
It clearly shows that the (amplitude-)equation \ref{eq:amplitudesys7} has an unique non-zero steady state of equal magnitude, i.e., $\overline{r}_{ss}=\sqrt{6}~(\approx 2.449)$. This steady state determines the amplitude of the limit cycle (cf.~\cite{Saha2019,Tri-rhythmic,Das2011CountinglcJKB}) and the stability of the limit cycle will defined the sign of the $\frac{d \dot{\overline{r}}}{d \overline{r}}|_{\overline{r}=\overline{r}_{ss}}$ (cf.~\cite{Saha2019,Tri-rhythmic,Das2011CountinglcJKB}). Here, $\frac{d \dot{\overline{r}}}{d \overline{r}}|_{\overline{r}=\sqrt{6}} < 0$ (for $\epsilon > 0$), and hence the limit cycle will be stable.}
\bibliographystyle{unsrt}
\bibliography{References}
\end{document}